# Lattice Quantum Geometry Controlling 118 K Multigap Superconductivity in Heavily Overdoped $CuBa_2Ca_3Cu_4O_{10+\delta}$

Gaetano Campi[1,2*], Massimiliano Catricalà[1], Giuseppe Chita[3], Luisa Barba[3], Luchuan Shi[4], Jianfa Zhao[4], Changing Jin[4*] and Antonio Bianconi[1,2]

[1] Institute of Crystallography, National Research Council, CNR, Via Salaria Km 29.3, 00015 Monterotondo Rome, Italy
[2] Rome International Center for Materials Science Superstripes RICMASS, Via dei Sabelli 119A, 00185 Rome, Italy
[3] Institute of Crystallography, National Research Council, CNR, Strada Statale 14 - Km163.5, Area Science Park, Basovizza 34149, Trieste, Italy
[4] Beijing National Laboratory for Condensed Matter Physics, Institute of Physics, Chinese Academy of Sciences, Beijing 100190, China.

| | |
|---|---|
| Gaetano Campi | https://orcid.org/0000-0001-9845-9394 |
| Massimiliano Catricalà | https://orcid.org/0009-0002-1426-1810 |
| Giuseppe Chita | https://orcid.org/0009-0006-3147-5414 |
| Luisa Barba | https://orcid.org/0000-0001-8832-7056 |
| Changing Jin | https://orcid.org/0000-0001-8097-9156 |
| Jianfa Zhao | https://orcid.org/0000-0002-7507-9441 |
| Luchuan Shi | https://orcid.org/0009-0000-1115-7984 |
| Antonio Bianconi | https://orcid.org/0000-0001-9795-3913 |

Email:

- Gaetano Campi * *email*: gaetano.campi@cnr.it
- Massimiliano Catricalà *email*: massimiliano.catricala@cnr.it
- Giuseppe Chita *email*: giuseppe.chita@cnr.it
- Luisa Barba *email*: luisa.barba@cnr.it
- Changing Jin* *email*: Jin@iphy.ac.cn
- Jianfa Zhao email: zhaojf@iphy.ac.cn
- Luchuan Shi email: shiluchuan15@iphy.ac.cn
- Antonio Bianconi * *email*: antonio.bianconi@ricmass.eu

* to whom correspondence should be addressed:
email: gaetano.campi@cnr.it
email: Jin@iphy.ac.cn
email: antonio.bianconi@ricmass.eu

## Abstract

Synchrotron X-ray diffraction has been used to study the thermal structure evolution in $CuBa_2Ca_3Cu_4O_{10+\delta}$ (Cu1234), a superconductor which exhibits a high critical temperature ($T_C \approx 118$ K), high critical current density and large upper critical magnetic field. The lattice geometry at nanoscale of this cuprate belongs to the class of natural heterostructures at atomic limit like the artificial high $T_c$ superlattices made of interface space charge in Mott insulator units intercalated by metal units. Temperature-dependent lattice parameters reveal a distinct lattice anomaly at $T_C$ characterized by a drop of the c-axis and in plane Cu-O negative thermal expansion below $T_C$. These results are consistent with multigap scenario and complex networks of multiscale configurations controlling macroscopic superconducting functions in complex perovskites. In the multigap scenario the lattice reorganization associated with the chemical potential changes could be assigned to the opening of multiple superconducting gaps in different points of the electron momentum space. Evidence for oxygen diffusion is observed at temperatures above 250 K. We construct a phase diagram correlating temperature, the c/a axis ratio, and in plane Cu-O strain, identifying regions associated with gaps opening and oxygen diffusion. These findings provide new insights into how lattice geometry controls superconductivity to inform the material design of advanced nanoscale superconducting artificial quantum heterostructures.

## Introduction

High-temperature superconductivity has been a subject within materials science focusing on the synthesis of new materials with improved superconducting properties. The field has captivated researchers due to underlying mesoscopic complex quantum matter and potential for groundbreaking applications. In these last years, complex nanostructured heavily overdoped cuprate perovskites, grown under high oxygen pressure and high temperature conditions, have emerged as a cornerstone of research for high critical temperature superconductors [1-10]. The $CuBa_2Ca_3Cu_4O_{10+\delta}$ (Cu1234) superconducting compound, belonging to the $CuBa_2Ca_{n-1}Cu_nO_{2n+2+\delta}$ homologous series, exhibits a critical temperature, $T_C$, exceeding 110 K at ambient pressure, comparable to Hg-based, which holds the highest $T_C$ among cuprates [11-14]. Furthermore, Cu1234 demonstrates superior critical current densities ($J_C$) at liquid nitrogen temperature, outperforming Bi-based superconductors and rivaling the widely used $YBa_2Cu_3O_{7-\delta}$ (YBCO) [15-24]. These cuprate superconductors are renowned for hosting emergent electronic phenomena, including enhanced superconductivity in the presence of multiple superconducting gaps [25-40]. It has been pointed out [32] that variations in the chemical potential at the superconducting transition temperature, $T_C$, resulting from the opening of superconducting gaps, become significant primarily in the case of multiband superconductivity, particularly when one of the Fermi surfaces lies near the edge of a conduction band. In such scenarios, the chemical potential experiences substantial changes between the normal and superconducting phases. This effect is especially pronounced when the chemical potential is tuned close to the band edge of a newly emerging band in the crossover regime. Here, the relative change in chemical potential exceeds that expected in the standard BCS regime, where such variation is typically negligible. This leads to a pronounced lattice anomaly driven by charge redistribution across bands with differing superconducting gaps.

Recent studies have underscored the pivotal role of lattice complexity in determining their superconducting properties [41-65]. This complexity encompasses lattice distortions proposed by Muller [47-55] and Goodenough [56-62], validated by experimental methods probing the local structure [63-65], oxygen interstitials rearrangements [42-46], strain in $CuO_2$ 2D layers, [66-68] and negative thermal expansion around $T_C$ [69-76]. Such lattice complexity is closely linked to nanoscale electronic phase separation [77-79]. While evolution of lattice complexity with temperature such as dopant oxygen interstitials diffusion has been extensively studied in systems like $YBa_2Cu_3O_{6+\delta}$ [42,43], $La_2CuO_{4+y}$ [44, 45], $HgBa_2CuO_{4+y}$ [46], analogous investigations on Cu1234 are lacking.

In this work, we explore the temperature-dependent lattice geometry of polycrystalline Cu1234 using synchrotron X-Ray Diffraction (XRD) across a wide temperature range. Our findings reveal a distinct lattice anomaly at $T_C$, characterized by a collapse of the c-axis and negative thermal expansion in the Cu-O plane. Additionally, we observe lattice fluctuations consistent with oxygen atom diffusion occurring above 250 K. By constructing a comprehensive phase diagram correlating temperature, lattice anisotropy and strain, this study sheds light on how lattice reorganization drives superconducting behavior in heavily overdoped cuprates.

A key aspect of our interpretation involves the layered structure of Cu1234, which alternates between metallic and Mott insulating units. This natural architecture resembles artificial nanoscale heterostructures, akin to Mott Insulator-Metal Interface (MIMI) systems [81-87]. Consequently, these natural and artificial superlattices provide a unique platform to investigate the interplay between distinct electronic states within $CuO_2$ planes, with potential implications for enhanced superconductivity. We focus on the experimental characterization of lattice quantum geometry and its changes at $T_c$ with the aim to present robust structural evidence of lattice distortions accompanying superconductivity.

**Results and discussion**

We have studied the temperature evolution of the polycrystalline Cu1234 synthesized under high oxygen pressure and high temperature conditions as described in [15]. The overall crystal structure of Cu1234 is typically tetragonal, in the P4/mmm space group. Rietveld refinement of Synchrotron XRD patterns collected on the XRD1 beamline at ELETTRA [88] on Cu1234 at different temperatures has been performed by using Expo2 [89] (see Methods). The XRD powder patterns collected with X-Ray wavelength of 0.7 Å, alongside the Rietveld best fitted lines, at T = 299 K and at T = 92 K are shown in Figure 1a. We find average volumetric thermal expansion coefficients $\Delta V/V\Delta T$ of $2.2\times10^{-5}$ $K^{-1}$ and $4.3\times10^{-5}$ $K^{-1}$ during the cooling and heating ramp, respectively. The doubling of this coefficient in the heating ramp could be ascribed to the interstitial oxygen rearrangement, as discussed ahead. The refined chemical composition of Cu1234, agreed with data determined through neutron powder diffraction [19]. More details on the refined structural parameters at 299 K and at T=92 K are listed in Table 1.

Cu1234 is a complex, layered perovskite-like structure composed of alternating Mott insulator [$Ca_3Cu_4O_8$] and metallic units [$Ba_2CuO_{4-y}$] with average valence state of copper Cu (+2.29) [19] establishing Cu1234 as a heavily overdoped layered cuprate superconductor [1-10]. The charge local redistribution has been investigated by Jarlborg et al. [76-79] in La-based and Hg-based cuprates revealing that dopant-induced charges in similar systems (e.g., $HgBa_2CuO_{4+\delta}$) are predominantly localized within the [$Ba_2CuO_{4-y}$] metallic layers, rather than uniformly distributed. This localization leads to the emergence of nanoscale phase separation forming a natural heterostructure at atomic limit [29] which is similar to Artificial High $T_c$ Superlattices (AHTS) fostering the amplification of superconductivity by Fano-Feshbach resonance in multi-gap systems with relevant Rashba spin-orbit-coupling at the interface [81-87].

In our Cu1234 sample, the metallic [$Ba_2CuO_{4-y}$] layers exhibit an unusual compressed octahedral coordination, differentiating them from conventional cuprates. This compression lifts the $3d_{z^2-r^2}$ orbital above the $3d_{x^2-y^2}$ orbital, altering the electronic hierarchy near the Fermi level in unconventional highly doped $Ba_2CuO_{4-y}$ [7-10, 93-101]. The charge disproportionation localized at the basal $CuO_2$ plane is shown by unusual, compressed Cu apical oxygen distances, significantly affecting orbital hybridization and promoting multiband superconductivity. At room temperature, the estimated Cu–O bond lengths are Cu1-O1=1.928 Å in-plane and Cu1-O2=1.768 Å along c-axis. This inverted bond-length hierarchy demonstrates that the $CuO_6$ octahedra are compressed like in $Ba_2CuO_{4-y}$ contrasting sharply with standard cuprates, where the Cu-O in-plane bond is typically shorter than the Cu-O bond along the c-axis.

As illustrated in Figure 1b, Ba1 O2 Cu1 O1 sites constitute the defective overdoped [$Ba_2CuO_{4-y}$] normal metal (N) units with thickness W=4.33 Å (blue thin units). Meanwhile Cu2, Cu3, Ca1, Ca2, O3, O4 form the tick superconducting (S) units, composed of a modulation-doped stoichiometric Mott insulator [$Ca_3Cu_4O_8$] with thickness L=13.61 Å, at low temperatures.

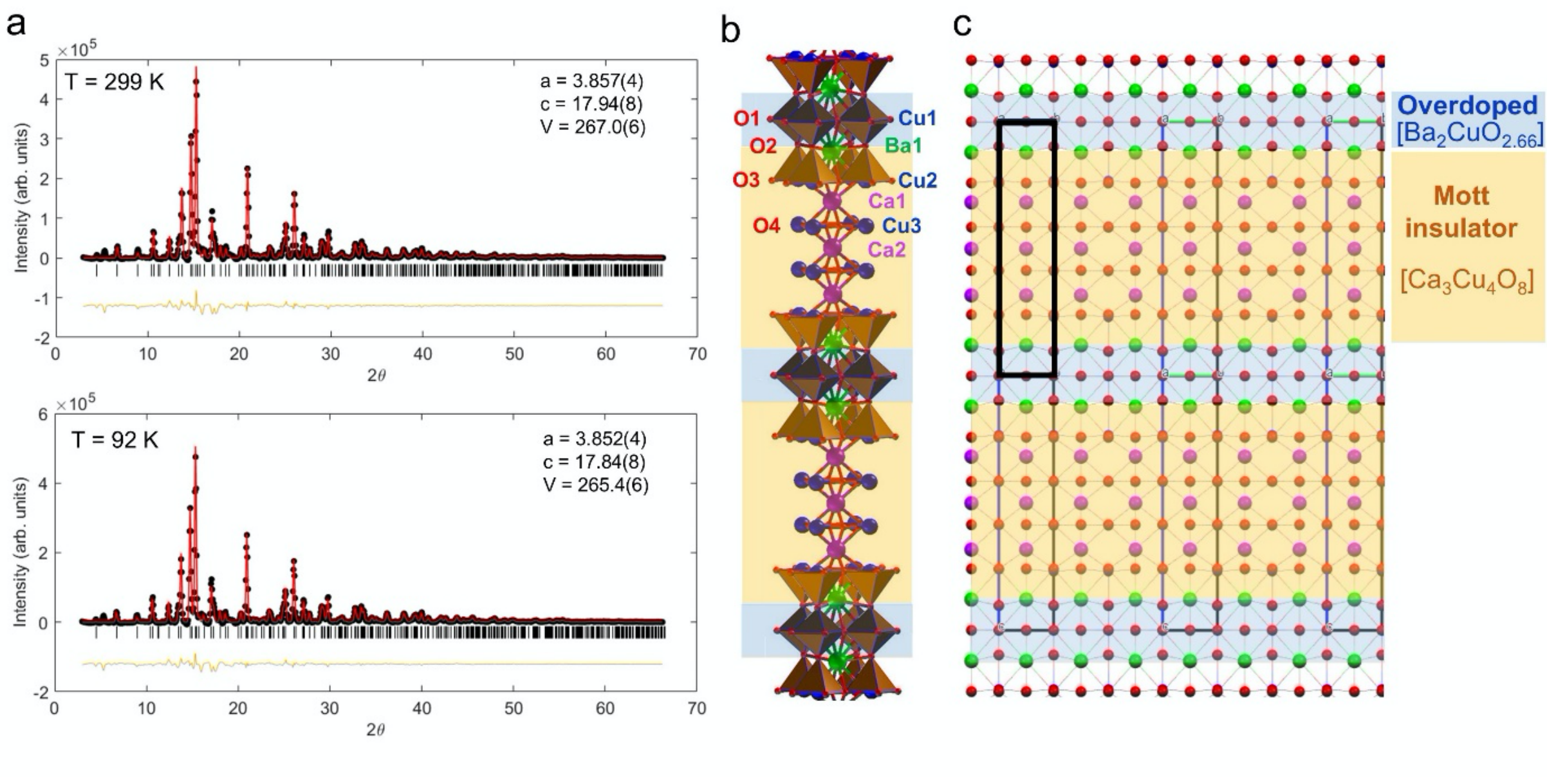

**Figure 1** (a) Rietveld refinement of the Cu1234 XRD patterns collected at T=299 K and T=92 K during a cooling ramp, using synchrotron radiation at ELETTRA (see methods). (b) Unit cells of Cu1234 along the c-direction with indicated atoms in the asymmetric unit. (c) View of crystalline packing highlighting the layered structure, recalling MIMI artificial nanoscale heterostructures at atomic limit, as described in the text. The blue light and orange strips represent the overdoped [$Ba_2CuO_{2+\delta}$] and hole doped Mott insulator [$Ca_3Cu_4O_8$] layers. The thick rectangle indicates the unit cell.

| T = 299 K | | | | | | T = 92 K | | | | |
|---|---|---|---|---|---|---|---|---|---|---|
| atom | x | y | z | $U_{iso}$(Å$^2$) | n | x | y | z | $U_{iso}$(Å$^2$) | n |
| **Ba1** | 0.5 | 0.5 | 0.1211(2) | 0.0213(14) | 1.000 | 0.5 | 0.5 | 0.1211(2) | 0.0188(10) | 1.000 |
| **Ca1** | 0.5 | 0.5 | 0.3162(6) | 0.011(3) | 1.000 | 0.5 | 0.5 | 0.3172(6) | 0.013(2) | 1.000 |
| **Ca2** | 0.5 | 0.5 | 0.5000 | 0.021(4) | 1.000 | 0.5 | 0.5 | 0.5000 | 0.018(3) | 1.000 |
| **Cu1** | 0.0 | 0.0 | 0.0000 | 0.078(6) | 0.94(3) | 0.0 | 0.0 | 0.0000 | 0.068(5) | 0.94(3) |
| **Cu2** | 0.0 | 0.0 | 0.2309(3) | 0.0085(16) | 1.000 | 0.0 | 0.0 | 0.2314(3) | 0.0090(13) | 1.000 |
| **Cu3** | 0.0 | 0.0 | 0.4115(3) | 0.0090(18) | 1.000 | 0.0 | 0.0 | 0.4116(3) | 0.0086(14) | 1.000 |
| **O1** | 0.0 | 0.5 | 0.0000 | 0.13(5) | 0.55(3) | 0.0 | 0.5 | 0.0000 | 0.13(5) | 0.55(3) |
| **O2** | 0.0 | 0.0 | 0.099(2) | 0.018(10) | 0.78(2) | 0.0 | 0.0 | 0.097(2) | 0.013(8) | 0.78(2) |
| **O3** | 0.0 | 0.5 | 0.2422(10) | 0.000(4) | 1.000 | 0.0 | 0.5 | 0.2412(10) | 0.001(4) | 1.000 |
| **O4** | 0.0 | 0.5 | 0.4145(11) | 0.014(6) | 1.000 | 0.0 | 0.5 | 0.4149(11) | 0.008(5) | 1.000 |

**Table 1** Fractional coordinates, x, y, z, isotropic Debye-Waller factor, $U_{iso}$, and occupancy, n, for each atom in the asymmetric unit of Cu1234 at room temperature, 299 K and after the first cooling cycle at T=92 K. We note the tendency to lower $U_{iso}$ factors at lower temperatures, as expected, except for the O1 atoms on the basal $CuO_2$ planes and O3. $R_p$ and $R_{wp}$ values are 6.33%, 7.64% at 299 K and 6.32%, 7.62% at 92 K.

The geometrical parameter characterizing the MIMI heterostructures superconducting performance is given by L/d [81-84] where L=d-W is the thickness of the metallic overdoped layer and d is the repeating units (c-axis). In this Cu1234 sample we have L/d=0.75, that is a value falling in the predicted range for the L/d values in proximity of the top of the superconducting dome [81-87]. The structure recalling the AHTS superconducting heterostructures at nanoscale [81-87] is depicted in Figure 1c.

Figure 2 illustrates the temperature-dependent structural evolution of the polycrystalline powder during cooling and heating cycles. Figure 2a shows colormaps of XRD intensity as a function of temperature and d-spacing for the 001 and 200 reflections, measured using an X-ray wavelength of 1.4089 Å. Gaussian fitting of these peaks across all temperatures allowed extraction of temperature evolution of unit cell parameters ***a*** and ***c***, as well as peak widths $\Delta d_{200}$ and $\Delta d_{001}$. Figure 2b presents the temperature-dependent variations of ***a*** and ***c***, normalized to the superconducting critical temperature ($T_C$). Three regimes are identified. In regime (1), below $T_C$, the ***c***-axis sharply collapses from 17.82 Å at $T_C$ to 17.70 Å at 92 K, accompanied by negative thermal expansion of the ***a***-axis. This behavior reflects lattice distortions, likely driven by anisotropic vibrations and $CuO_6$ octahedral rotations, as proposed by Purans et al., [73] where rotations or distortions of $CuO_6$ octahedra in the perovskite-type $ScF_3$ structure are driven by anisotropic thermal vibrations. In regime (2) between $T_C$ and the oxygen ordering temperature ($T_O$~250 K), the *c*-axis exhibits contraction and expansion rates ($5.16\times10^{-5}$ $K^{-1}$) during cooling and heating, while the ***a***-axis expands with rate $0.71\times10^{-5}$ $K^{-1}$. In regime (3), above $T_O$, the heating ramp causes rapid expansion of the ***c***-axis from 17.89 Å at $T_O$ to 17.94 Å at 299 K, contrasting with its stability during cooling.

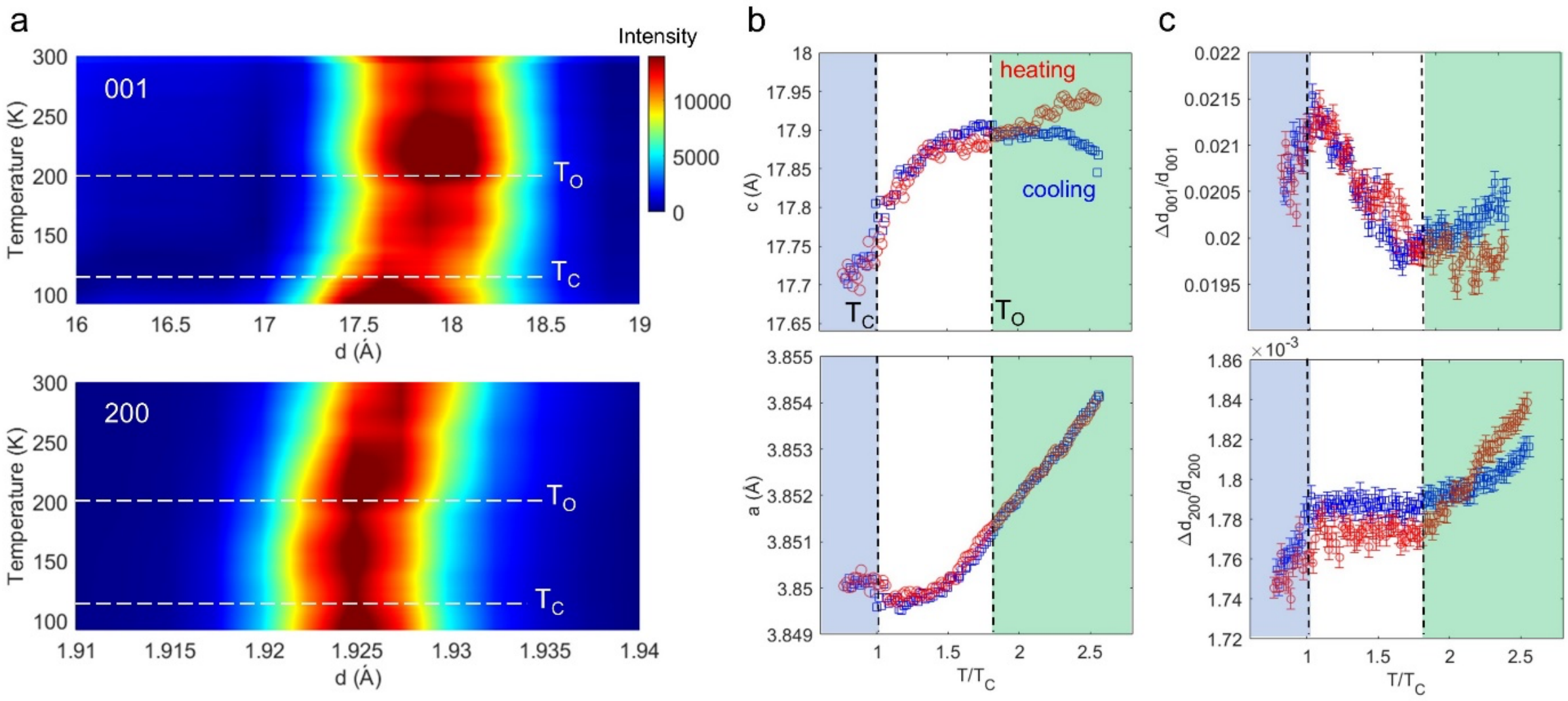

**Figure 2** (a) Colormaps of temperature dependence of 001 and 200 reflections. The superconducting and oxygen ordering temperatures $T_C$ and $T_O$ are indicated (white dashed lines). Temperature dependence of (b) unit cell parameters, c and a, and (c) fluctuations $\Delta d_{001}/d_{001}$ and (f) $\Delta d_{002}/d_{002}$. We observe a lattice anomaly at $T_C$ where the c-axis sharply drops from 17.82 to 17.70 Å in the superconducting phase and the a-axis sharply increases from 3.8495 Å to 3.850 Å. The broadening of XRD peaks 001 and 200 at $T_C$, in panel (c) confirm the different regimes in the structural temperature evolution and the lattice anomaly at $T_C$.

This cusp is attributed to oxygen rearrangements along the *c*-axis. Figure 2c highlights fluctuations in lattice spacing, $\Delta d_{200}/d_{200}$ and $\Delta d_{001}/d_{001}$, across these regimes. Below $T_C$, fluctuations decrease sharply, stabilizing the superconducting state. Between $T_C$ and $T_O$, in-plane fluctuations remain constant while out-of-plane fluctuations reflect stronger contraction during cooling. Above $T_O$, both in-plane and out-of-plane fluctuations follow temperature trends, indicating lattice stabilization due to oxygen reorganization. These observations underscore how structural anisotropy and lattice distortions are closely tied to superconductivity stabilization [70-74].

Since the superconducting properties of Cu1234 primarily originate from the $CuO_2$ planes, which are the central structural feature governing the transport properties of cuprate superconductors, we have characterized the Cu-O planar structure through the calculation of the in-plane strain. The strain is defined as $\varepsilon=2*100*(Cu\text{-}O_{eq}-Cu\text{-}O_{obs})/Cu\text{-}O_{eq}$, where $Cu\text{-}O_{eq}=1.97$ Å represents the Cu-O bond distance in equilibrium conditions, as determined in $Cu^{2+}$ ions in solution, and $Cu\text{-}O_{obs}$ is the observed bond distance Cu1-O1 in the basal plane, under our experimental conditions [65-67]. The factor of 100 is included to express the strain as a percentage. We define a critical strain, $\varepsilon_c=4.57\%$, corresponding to its value at the superconducting critical temperature $T_C$. Using this definition, a comprehensive phase diagram has been constructed (see Figure 3), plotting the crystallographic axis ratio, c/a, and the normalized temperature as functions of the normalized strain $\varepsilon/\varepsilon_c$.

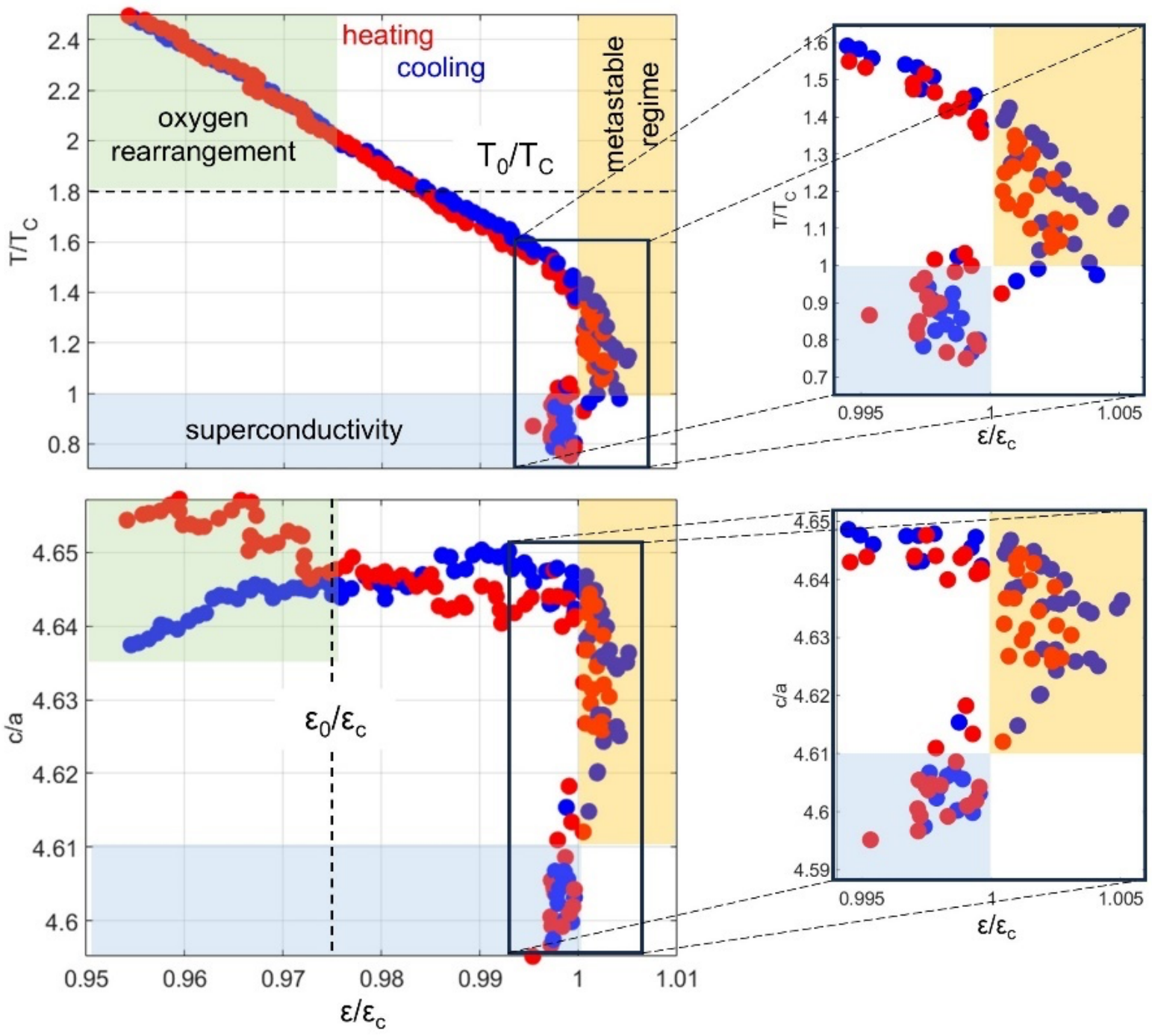

**Figure 3** (upper panel) Normalized temperature, $T/T_C$, and axis ratio, c/a, as a function of normalized strain, $\varepsilon/\varepsilon_c$. The superconducting phase (blue light area) occurs for $T<T_C$ and $\varepsilon<\varepsilon_c$ where $T_C$ = 118 K and $\varepsilon_c$ = 4.57%. The oxygen rearrangement occurs for $T>T_O$ and $\varepsilon<\varepsilon_O$. When $T>T_C$ and $\varepsilon>\varepsilon_c$ the system undergoes to a metastable phase (yellow area). Finally, for $\varepsilon<\varepsilon_c$ and $T>T_C$ we observe a decreasing strain. The structural phase transition at $T_C$ is highlighted in the magnified panels on the right.

Here we can see the distinct regimes previously described and indicated by colored areas, linked to the interplay between superconductivity and structural evolution. In the superconducting regime $T<T_C$ and $\varepsilon<\varepsilon_C$ (blue light area), the $CuO_2$ planar structure undergoes a stabilization process, reflecting the reduction in strain. Additionally, a metastable phase is identified for $\varepsilon>\varepsilon_c$ during cooling below $T_m$=165 K, as highlighted in the orange-shaded region in Figure 3. This phase likely arises from localized structural instabilities within the perovskite Cu-O planar bonds, reflecting their critical role of lattice dynamics in high $T_c$ superconductivity.

In Figure 4 we show the Cu(1)-O(1) bond distances derived from Rietveld refinement of synchrotron X-ray diffraction data collected using incident wavelength of 0.6999 Å. Due to limitations in angular resolution imposed by the PILATUS 2000 detector (pixel size: 172 x 172 μm²), the refined Cu(1)-O(1) bond distance exhibited larger fluctuations in comparison with Cu(1)-O(1) distances obtained through direct fitting of the [2 0 0] Bragg reflection, where the Cu(1)-O(1) distance corresponds to half the a-axis lattice parameter. This lower-energy dataset yielded smoother trends thanks to improved low-angle resolution, although it exhibited less distinct variations in its temperature-driven evolution, as shown in previous Fig. 3. Anyway, the convergence of both methods underscores the robustness of the observed structural evolution.

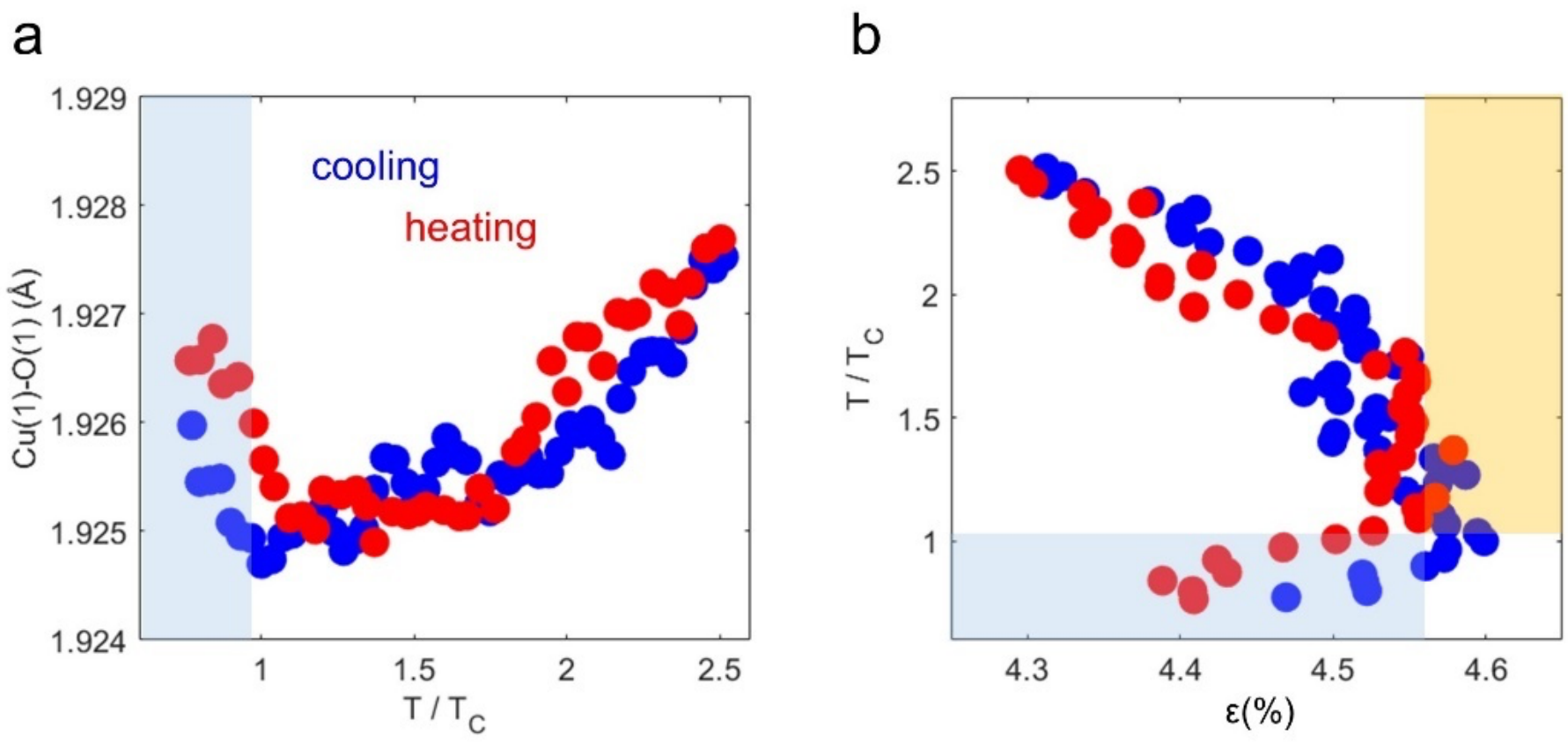

**Figure 4**: (a) Cu(1)-O(1) bond distances derived from Rietveld refinement of synchrotron X-ray diffraction data collected at incident wavelengths of 0.6999 Å during cooling (blue circles) and heating (red circles) thermal rampes. (b) Normalized temperature, $T/T_C$, as a function of strain, ε. The light blue areas indicate the superconducting state, while the orange zones are associated with the metastable phase.

Conversely, in the regime $T>T_O$ and $\varepsilon<\varepsilon_O$, (green area), where $\varepsilon_O$=4.49% corresponds to the strain at $T=T_O$, structural fluctuations become irreversible, as shown in Figure 2b by the c-axis temperature evolution. We report the Debye–Waller factors, U, of oxygen atoms to complement our analysis of geometrical fluctuations via Bragg reflection widths (see Figure 2). We have analyzed the temperature-dependent behavior of U at distinct oxygen sites. We have observed that U(O2), U(O3) and U(O4) exhibit a drop below $T_C$, indicating reduced atomic displacements and supporting the stabilization of the superconducting state. Between $T_C$ and $T_O$≈250 K, U(O2) and U(O3) remain nearly constant, while U(O4) follows the temperature evolution. Above $T_O$, all three U values increase upon heating and decrease upon cooling, closely mirroring c-axis trends and revealing irreversible reorganization of oxygen motion (see Figure 2b). Meanwhile, U(O1) remains consistently high and temperature-independent, pointing to persistent larger positional disorder. These findings look to indicate $T_O$ as a temperature for oxygen diffusion in line with phenomena observed in other cuprate superconductors [42-46]. In particular thermal treatment have shown the onset of interstitial oxygen rearrangement at temperatures larger than 200 K in $La_2CuO_{4+y}$ [45].

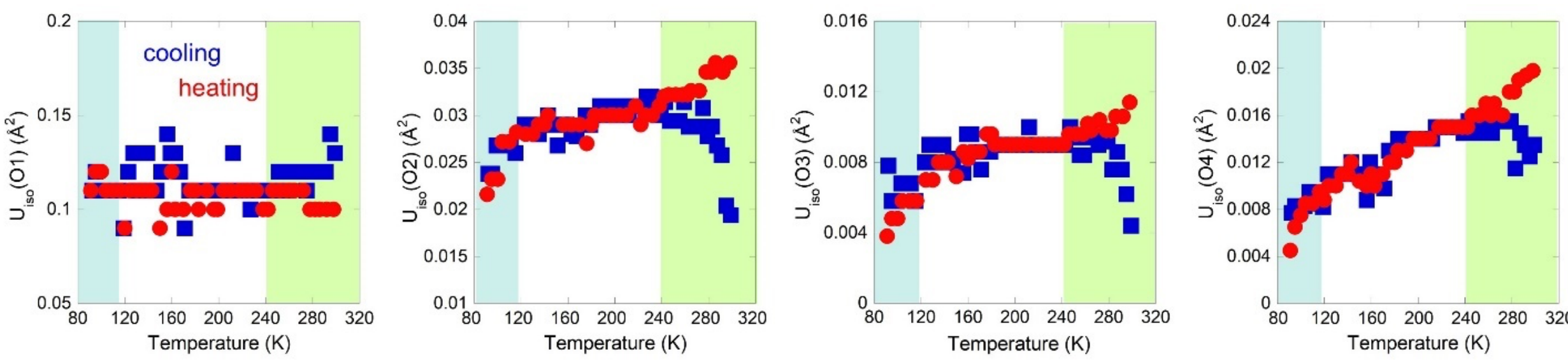

**Figure: 5** Debye Waller factors extracted from Rietveld refinement for oxygen in O1, O2, O3 and O4 sites. The light blue areas indicate the superconducting state, while the light green zones are associated with the oxygen diffusion temperature range.

Standard BCS single-gap theory does not predict lattice rearrangements at $T_C$. In Cu1234, however, the described sharp c-axis contraction of 0.12 Å (from 17.82 Å to 17.70 Å) and negative in-plane thermal expansion below $T_C$, represent anomalies beyond the scope of conventional BCS behavior. We interpret this as evidence for multi-gap superconductivity, arising from the heterostructure's layered architecture comprising metallic [$Ba_2CuO_{4-y}$] and Mott insulating [$Ca_3Cu_4O_8$] planes, which foster charge disproportionation and multiband effects supported by multiple Fermi surfaces, measured by ARPES experiments in Hg-based cuprates [76–78]. Such anomalies may also reflect significant chemical potential shifts near the 3D–2D electronic topological transition, where multiband shape resonances [32] amplify superconducting gap interactions across distinct bands. These combined factors support the proposed multi-gap superconductivity scenario offering a more consistent explanation than a single-gap scenario.

We have integrated our experimental findings for Cu1234 and the AHTS heterostructures [80] into the general "$T_C$-strain-doping" phase diagram of cuprates [64], as shown in Figure 6. The Cu1234 superconductor, characterized by a strain ε=4.3 % at room temperature, doping δ = 0.29, and the high $T_C$ of 118 K [19] aligns with the strain-doping landscape of cuprates. Indeed, although this system resides in the overdoped regime its strain leads it in proximity to the top of the superconducting dome for all families of standard hole doped cuprate perovskites. In contrast, the artificial high $T_c$ superlattices (AHTS) composed of La-based cuprates alternating dopants Sr rich units and stoichiometric $CuO_2$ units exhibits a higher compressive strain ε=9% and a significantly lower $T_C$ of 43 K [65]. The elevated strain in MIMI shifts the system away from the ($\varepsilon_C$≈0.04 for $\delta_C$=0.16), suppressing $T_C$. Thus, our Cu1234 system shows how strain engineering provides a pathway to optimize superconductivity at ambient pressure in cuprate perovskites.

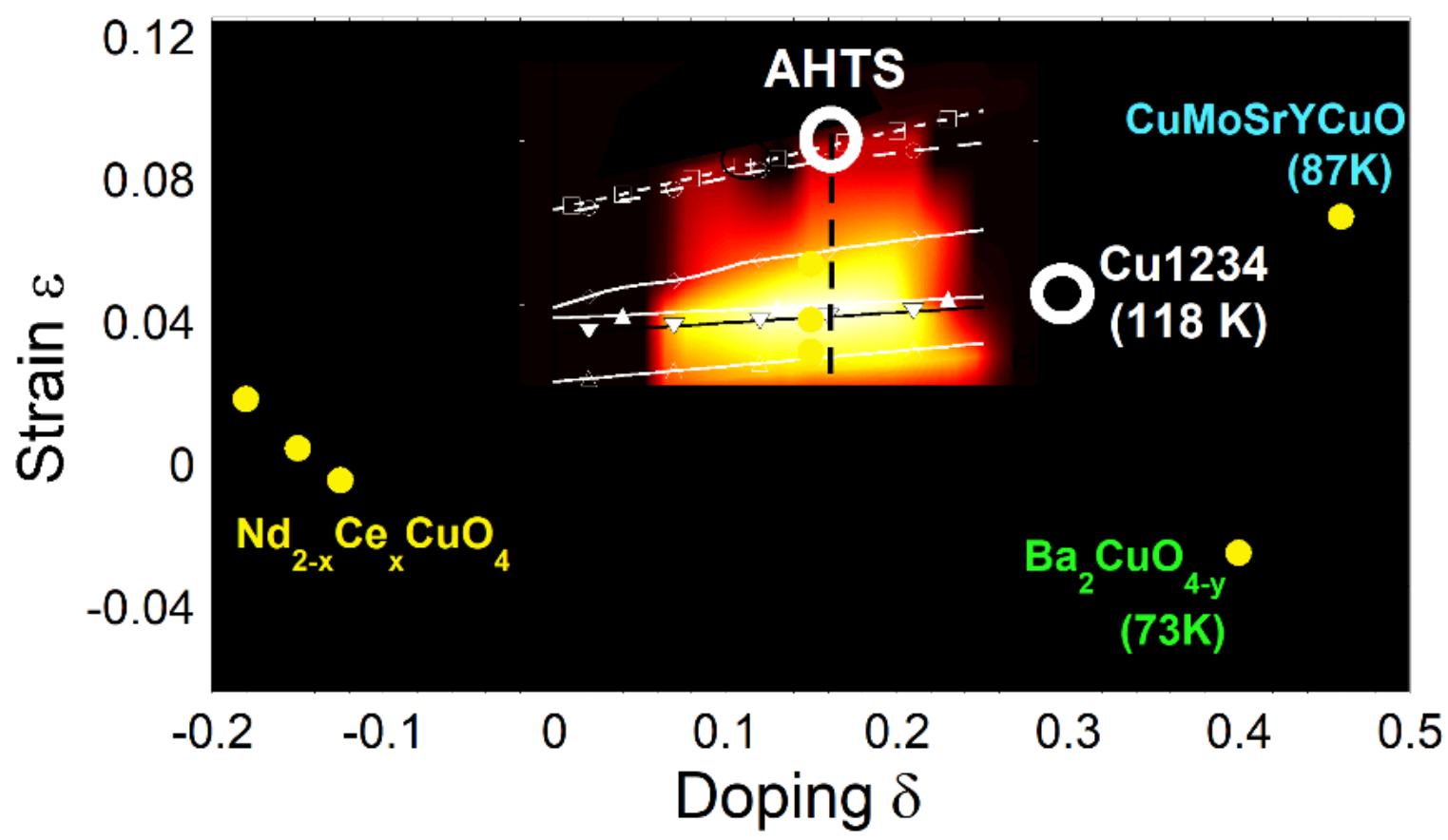

**Figure 6** General phase diagram for different cuprate perovskite families as a colormap of $T_C$ (from $T_c$ = 0 K, black, to $T_c$ ~ 135 K, through yellow to white) as a function of strain, ε, and doping, δ [64,65]. Small open white symbols represent hole doped cuprates. Maximum $T_C$ occurs at δ=0.16 and ε=4%. The yellow full circles represent the electron doped and highly overdoped cuprates [1-10, 89-96]. The thick white empty circles indicate the Cu1234 sample with average doping δ= 0.29 and strain 4% studied in this work compared with the La-based AHTS with $T_C$=43 K [81-86].

Conversely, the MIMI structure underscores the lower superconductivity under larger strain. This duality emphasizes the need to balance doping and strain to engineer high-performance superconductors. The lattice anomaly associated with the superconducting transition provides key insights into the behavior of cuprates, particularly in the context of the resonant multigap superconductivity with relevant electron-lattice interaction [81-87]. In Cu1234, the lattice anomaly observed at the superconducting critical temperature is generated by the variation of the chemical, potential as occurs in systems with multiple different gaps opening at $T_c$ with a charge redistribution due to relevant electron lattice interactions tuned by strain identified as a critical factor [66–68]. Specifically, the interplay between lattice distortions and electronic properties in Cu1234 is evident in two key phenomena below $T_C$: the expansion of the $CuO_2$ planes and the compression of the interlayer spacing along the c-axis. These structural adjustments point to a strong coupling between electrons and phonons in hot spots of the electronic k-space, as well as multiband superconductivity. This interplay is tuned by strain favoring anisotropic lattice distortions due to chemical potential changes at the opening of multiple superconducting gaps.

## Conclusions

Synchrotron X-ray diffraction measurements of Cu1234 reveal significant insights into the interplay between lattice geometry and superconductivity in this strongly overdoped high temperature cuprate. The structure of Cu1234 can be classified as a layered MIMI structure. This architecture, comprised of alternating hole doped Mott insulator [$Ca_3Cu_4O_8$] units and metallic [$Ba_2CuO_{2+\delta}$] layers, is crucial to its unique properties. The temperature-dependent analysis of the unit cell parameters, ***a*** and ***c***, and the in-plane strain unveils distinct structural changes at the superconducting critical temperature, $T_C$~118 K. Specifically, we observe anisotropic thermal expansion below $T_C$, with a negative thermal expansion along the ***a***-axis coupled with a contraction along the ***c***-axis. Furthermore, fluctuations in the d-spacings, $\Delta d_{001}/d_{001}$ and $\Delta d_{200}/d_{200}$, and oxygen Debye-Waller factors exhibit clear differences across the different temperature regimes distinguished by superconducting and oxygen diffusion temperatures $T_C$ and $T_O$, akin to observations in several cuprate high temperature superconductors. Analyzing the in-plane *strain*, derived from the Cu-O basic planar structure, we construct a phase diagram correlating normalized temperature ($T/T_C$) and axis ratio (c/a) as a function of normalized strain ($\varepsilon/\varepsilon_c$). This diagram identifies a clear structural transition between $0.8<T/T_c<1.3$ with a compression of the c-axis and a sharp expansion of the a-axis at $T_C$ with a decreasing strain which is maximum around $T/T_c \approx 1.3$. The observed lattice anomaly at $T_c$ is intrinsically linked to the opening of multiple superconducting gaps. Furthermore, the compressed local Cu1 octahedron indicates that the charge introduced by doping is localized on the strongly overdoped [$Ba_2CuO_{4-y}$] layers. This charge localization provides a nanoscale electronic phase separation within is likely a key factor in the enhancement of superconducting properties. The metallic [$Ba_2CuO_{4-y}$] layers (of thickness W =

4.33 Å) play the role of overdoped metallic units, while the [$Ca_3Cu_4O_8$] blocks (of thickness L=13.25 Å) host the confined superconducting interface space charge forming a MIMI. The nanoscale high $T_c$ superlattice of quantum wells with period d=1.758 nm is characterized by the geometry ratio L/d =0.75 positioned near the top of the superconducting dome observed in artificial MIMI systems [81–86]. The Cu1234 structure achieves optimal strain-doping synergy, enabling its high $T_C$~118 K. This natural MIMI configuration creates distinct $CuO_2$ planes: chemically overdoped metallic layers and chemically undoped layers. This underscores a fundamental departure from conventional cuprate models, where $CuO_6$ compressed local octahedron and heavily overdoped hole carriers [91-96] redefine the interplay between charge disproportionation and orbital hybridization, leading to multigap high-$T_c$ superconducting mechanisms observed in artificial high-$T_C$ superlattices [81–86]. The resulting charge and lattice phase separation as predicted by Jarlborg et al. [77–80] confined in Cu1234 by XANES spectroscopy [19] drives an emergent quantum electronic phase with nanoscale quantum size effects, proximity to a Lifshitz electronic topological transition in presence of spin orbit coupling at the interface. Such interfacial heterogeneity enhances quantum coherence in interface superconductivity in the stoichiometric units, a mechanism predicted by the BPV theory and verified experimentally in artificial high-$T_C$ superlattices [81–86].

By integrating these findings into the universal "$T_C$-doping-strain" phase diagram, shown in Fig.4 where we observe that the high average p=0.29 hole/Cu site doping and the strain ε=4.2% places Cu1234 at the optimal strain for the $T_C$ maximum [66].

Our results highlight the importance of nanoscale phase separation [49-53,97-100] and the geometry of nanoscale structural complexity in governing the superconducting properties of Cu1234. Like in AHTS [29-33,81-87,101] where high $T_c$ is controlled by the optimal lattice strain and geometry ratio 0.6<L/d<0.75 in the special case where the metal units are heavily overdoped copper-oxide planes. Finally, our crystallographic and lattice distortion analyses offer a solid experimental foundation for characterizing the structural evolution of Cu1234 across the superconducting transition which was observed previously in other cuprates [36] using X-ray absorption spectroscopy with synchrotron radiation, a fast and local probe of nanoscale structure of complex matter [63-65,102].

**Methods**

The X-ray diffraction (XRD) measurements for the powdered Cu1234 sample were conducted on the XRD1 beamline at the ELETTRA synchrotron radiation facility, Trieste, employing high-resolution transmission geometry [88]. Data acquisition spanned two thermal cycles, each consisting of a cooling ramp followed by a heating ramp. In the first thermal cycle, measurements were performed using a photon wavelength of 0.7 Å, which provided higher resolution and was therefore employed for structural Rietveld refinement analysis. During the second cycle, a wavelength of 1.4809 Å was used, optimized for studying the temperature evolution of structural parameters through more precise fitting of the 001 and 200 reflections. The Pilatus2M detector was positioned 86 mm from the sample. The sample temperature was systematically varied between 90 K and 300 K. At each temperature set point, the system was allowed to equilibrate until the temperature gradient within the sample was less than 0.1 K, ensuring reliable measurements. All acquired diffraction images were processed using the FIT2D software suite. Structural analyses were performed using the Expo2 Rietveld refinement program [89], while additional temperature-dependent parameter evaluations were conducted with custom MATLAB routines developed in-house.

**Acknowledgment**

We thank the Superstripes onlus and the CNR project DCM.AD006.562 “Functional Disorder in Materials and Complex Systems” for supporting this work. We thank the Institute of Physics, Chinese Academy of Sciences (IOPACS) for financial support.